# Thermal Wave Induced Edge Electrical Field of Pyroelectric: Spatial Pattern Mapping and Effect of Ambient Conditions.

A.V. Butenko, V. Sandomirsky, G. Chaniel, B. Shapiro and Y. Schlesinger

Department of Physics, Bar-Ilan University, Ramat-Gan 52900, Israel

A. Jarov, Raicol Crystal LTD, P.O.Box 2753, Yehud, 56217, Israel

V.A. Sablikov, Institute of Radio Engineering and Electronics, Russian Academy of Sciences, Fryazino, 141190, Russia

Abstract.

We have recently analyzed theoretically the main characteristics of the edge depolarizing electric field (EDEF), in the vicinity of a *non-polar face* of a pyroelectric. In this work we measured and characterized the EDEF, excited by a harmonical thermal wave. We present here experimental results obtained on a pyroelectric crystal $LiTaO_3$, confirming our theoretical predictions. We present the theoretical analysis and description of the thermal wave and the induced harmonically varying EDEF. The calculations assume an equivalent circuit of a pyroelectric capacitive current source. The measured magnitude of the EDEF and its spatial variation agree well with the theoretical model. The effect of the air pressure at the pyroelectric/air interface, on the EDEF, was determined in the interval $10^3 - 10^{-6}$ torr. We found that EDEF increases significantly with decreasing air pressure, presumably due to diminishing of adsorption screening at the polar faces. Teflon plates, covering the polar faces, prevent accumulation of screening charged particles, resulting in a drastic increase of EDEF.

Introduction

The transient EDEF, appearing after applying a temperature step ($\Delta T$), has been calculated and published recently [1]. We pointed out the potential of developing a broad spectrum of new devices and technologies, depending on the spatial distribution and magnitude of the EDEF. This makes it mandatory to corroborate experimentally the theoretically predicted properties of EDEF.

The pyroelectric (PE) with EDEF is depicted schematically in Fig. 1a, exhibiting clearly the typical pattern of a macroscopic electric dipole field. It follows then, that the EDEF extends in space over a distance of the order of the dipole length. Also, the characteristic magnitude of EDEF is equal to the change of the polarization charge surface density

$$E \simeq \Delta\sigma = p\Delta T, \qquad (1)$$

where $p$ is the pyroelectric coefficient.

In a good PE, such as LiTaO$_3$ ($p \simeq 2\times 10^{-2} \mu C/cm^2 K$), with an area of polar faces of ~ 1 cm$^2$, and $W$ ~ 1 cm (Fig.1 b) a temperature change $\Delta T \simeq 50\,K$ (in air) results in a EDEF of $\simeq 10^4 - 10^5$ V/cm.

The study of the realistic EDEF involves elucidation of the various physical factors affecting the measured value of EDEF. Following an application of a temperature step pulse, the field relaxes mainly due to screening the polarization surface charges by charged particles adsorbed from the surrounding medium. A survey of up-to-date literature [2] reveals scarce or only qualitative reference to this process. Yamagushi [3] deals with the adsorption of some gases on the tourmaline surface, and, although rather interesting, it does not provide the detailed necessary information. Therefore, we undertook an investigation to study the strong influence of the ambient air pressure, on the EDEF, in the range of $10^{-6} \div 760$ torr.

The pyroelectric effect is of a transitory nature. Its relaxation creates a transient EDEF in the PE vicinity, accompanied by a displacement current which is proportional to EDEF. Thus measuring the displacement current allows determining the magnitude of EDEF. However, the pyroelectric effect relaxation, following the application of the temperature step $\Delta T$, is not easily controllable. To turn this relaxation into a well defined process, we have chosen to apply a harmonically varying temperature, allowing a relatively simple theoretical formulation.

The harmonical thermal wave, propagating through the PE, excites a periodically varying polarization wave, which in turn, creates the harmonical EDEF (HEDEF). In the next sections we present the solution of the heat transport, followed by the calculation of the electrical potential and electrical field for quasi-stationary HEDEF. Further, we describe an equivalent scheme of PE as a capacitive current source, serving as the basis for the interpretation of the measurement method. Finally, we present and discuss the experimental data and their interpretation.

2. Theoretical model.

2.1. The thermal wave.

Fig. 1b shows the PE-bar, attached to a thermoelectric element, harmonically varying the temperature at $y = -H$. The spatial and temporal temperature variation of the PE is described by the one-dimensional thermodiffusion equation for the PE-bar bounded by air

$$\frac{\partial T}{\partial t} = \chi \frac{\partial^2 T}{\partial y^2}; \quad t > 0; \quad -H < y \leq 0;$$

$$\frac{\partial T_a}{\partial t} = \chi_a \frac{\partial^2 T_a}{\partial y^2}; \quad t > 0; \quad 0 < y < \infty, \tag{2}$$

where $T(y, t)$ is *temperature change* of the PE, $\chi$ is its thermodiffusivity, $T_a(y, t)$ is the temperature variation of the surrounding air. The boundary conditions are

$$T(-H) = T_g \exp(-i\omega t)$$

$$\kappa \left.\frac{\partial T}{\partial y}\right|_0 = \kappa_a \left.\frac{\partial T_a}{\partial y}\right|_0 \tag{3}$$

$$T(0) = T_a(0); \quad T_a(\infty) = 0,$$

$T_g = T(-H, 0)$, $\omega = 2\pi f$ is the temperature alternation frequency, $\kappa$ and $\kappa_a$ are the thermoconductivities of PE and air, respectively. Since very low frequencies have been applied, it is convenient to use the period $\tau = 1/f$ instead of frequency. The solution of Eqs. (2) and (3) is the attenuated thermal wave

$$T(y,t) = T_g \exp(-i\omega t) \frac{\cosh(ky) - b\sinh(ky)}{\cosh(kH) + b\sinh(kH)}, \quad -H \leq y \leq 0, \tag{4}$$

$$k_0 = \sqrt{\frac{\pi}{\chi\tau}}; \quad k = (1-i)k_0; \quad b = \sqrt{\frac{\kappa_a c_a \rho_a}{\kappa c \rho}}, \tag{4.1}$$

where $c_a, c, \rho_a, \rho$ are the heat capacities and densities of air and PE, respectively; $k_0$ is the wave vector; $b$ is a characteristic of the thermal barrier at the PE/air interface.

For the comparison with experiment, the amplitude of the thermal wave is

$$\tilde{T}(y) = T_g \frac{\left[(\cosh\alpha - b\sinh\alpha)^2 \cos^2\alpha + (b\cosh\alpha - \sinh\alpha)^2 \sin^2\alpha\right]^{1/2}}{\left[(\cosh\beta + b\sinh\beta)^2 \cos^2\beta + (b\cosh\beta + \sinh\beta)^2 \sin^2\beta\right]^{1/2}}, \tag{5}$$

where $\alpha = k_0 y$, $\beta = k_0 H$.

The details of the derivation of the last formulae is given in Appendix 1.

Fig. 2a shows the 3D-graph of the attenuated thermal wave $T(y,t)/T_g$ inside the PE as a function of $y$ $(-H \leq y \leq 0;\ H = -0.7\,\text{cm})$ and of the phase $\delta = \omega t$. Fig. 2b displays the sections of the 3D-graph by planes of $\delta = const$ in the semi-period interval, $0 < \delta < \pi$ (in the interval $\pi < \delta < 2\pi$ the function $T/T_g$ is the same except for opposite sign), showing the momentary snapshots of the temperature distribution along PE. The values $b = 0.55$ and $\chi = 0.0115$ cm$^2$ are experimental results, $\tau = 100$ s is a typical period. It should be noticed that there are instants, when the sign of $T(y)$ changes along the sample.

At $k_0 H < 1$ (small $\tau$) the thermal wave is strongly attenuated within the PE, and the boundary conditions, Eq. (2), at $y = 0$ are $T(0) = 0$. The wave propagates as in a semi-infinite bar, and is unaffected by the thermal interface with air. This fact will be used later at the treatment of experiment.

To calculate the function $T(y, t)$ numerically, the two parameters, $b$ and $\chi$, appearing in Eqs. (4) or (5) have to be given explicitly. The values of the thermodiffusivity $\chi$ of LiTaO$_3$, used in the present experiments, as appearing in literature, vary rather significantly. Therefore, we decided to determine $\chi$ from our experiment. The thermodiffusivity is a characteristic of substance, hence, it does not depend on the surrounding medium. The quantity $b$, on the other hand, depends on the properties of the interface of the LiTaO$_3$/air thermal barrier. Thus, it depends on the ambient air pressure, which will affect the temperature profile in PE, and, by this, the HEDEF.

2.2. The electrical field HEDEF

The thermal wave along the $y$-axis creates, within the PE, a harmonically varying polarization change, $\Delta \vec{P}(y,t)$, in the $z$-axis direction (see Fig. 1b). Therefore, $\nabla \cdot \Delta \vec{P}(y,t) = 0$, hence, this wave does not generate a bound space charge. The surface polarization charges on faces $z = \pm L$, vary harmonically only.

To avoid possible misunderstanding, it should be remembered that the direction of the polarization in PE (oppositely to its magnitude) does not depend on temperature. Let the plane at $z = L$ possesses positive surface charge, and the face at $z = -L$ negative surface charge. Then, these polarization charges vary accordingly as

$$\sigma^+(L) = \sigma_0 + \Delta\sigma(y)\exp(-i\omega t) > 0;$$
$$\sigma^-(-L) = -[\sigma_0 + \Delta\sigma(y)\cdot\exp(-i\omega t)] < 0, \quad (6)$$

where

$$\Delta\sigma(y) = pT(y), \quad (7)$$

and $T(y)$ is given by Eq. (4). In equilibrium $\sigma_0 = 0$, or $const \simeq 0$. The varying surface charges at both faces, $\Delta\sigma(y)\exp(-i\omega t)$, create an alternating quasi-stationary DEF inside, and EDEF outside the PE. The HEDEF creates a displacement current in surrounding space. The entire PE is enwrapped by HEDEF (see Fig. 1a), however we are here interested only in HEDEF in the region $y > 0$, where this field has practical application importance and is studied here experimentally.

The PE bar has a width 2L, (-L< z < L), and a thickness 2W, (-W < x < W) Then, the quasi-stationary potential generated by the charges $\pm\Delta\sigma(y)$ is (see Appendix 2):

$$\Phi(x, y.z, t) = \Phi_1(x, y.z, t) + \Phi_2(x, y.z, t), \quad (8)$$

where

$$\Phi_1(x, y.z,t) = 2\int_0^W du \int_{-H}^0 dv \frac{-pT(v,t)}{\sqrt{(x-u)^2 + (y-v)^2 + (z+L)^2}}$$

$$\Phi_2(x, y.z,t) = +2\int_0^W du \int_{-H}^0 dv \frac{pT(v,t)}{\sqrt{(x-u)^2 + (y-v)^2 + (z-L)^2}} \qquad (8.1)$$

We assume here that the dielectric constant of PE, and that of the surrounding air are equal, $\varepsilon = 1$. This assumption overestimates HEDEF by a factor of $\approx \varepsilon$, but does not distort its spatial pattern.

We introduce the dimensionless variables, parameters, and normalized potential

$$\frac{v}{L} \to v; \frac{x}{L} \to x; \frac{y}{L} \to y; \frac{z}{L} \to z; \frac{H}{L} = h; \frac{W}{L} = w; \qquad (9.1)$$

$$\eta = k_0 L v \text{ (here } v \to \frac{v}{L}\text{); } \beta = k_0 L h; \delta = \omega t; \qquad (9.2)$$

$$U(x, y, z, t) = U_1 + U_2$$
$$U_1 = \frac{\Phi_1}{2pT(-H)L}; U_2 = \frac{\Phi_2}{2pT(-H)L} \qquad (9.3)$$

The explicit formulas for the potential are derived in Appendix 2. Fig. 3 illustrates the 3D-shape of the potential, $U(0,y,z)$, constructed using Eq.(9.3), at some arbitrary moment and parameter values, same as in Fig. 2. The potential is antisymmetric with respect to the plane $z = 0$. The potential vanishes at the plane $z = 0$, and $U \to 0$ at $x, y, z \to \infty$. It has a ridge and a valley along the directions $z = \pm 1$. The points of maxima and minima are located at $y = 0$. Fig. 3 shows that the potential decays over a distance of ~$L$.

The amplitude of the potential, $\tilde{U}$, increases with $\tau$ (Fig. 4a), and its maxima and minima become sharper (Fig. 4b). The maximum of the dimensionless potential is $\simeq 1$. The characteristic value of the real potential, obtained by applying Eq. (9.3), is $2pT_gL$. In LiTaO$_3$ ($p = 2 \times 10^{-8}$ C/cm$^2$K) at an amplitude of $\tilde{T}_g = 1$ K and $L = 1$ cm this value is $\simeq 4 \times 10^4$ V.

The dimensionless electrical field was calculated using the potential given by Eq. (9.3) and using the definition

$$\vec{E} = \nabla U(x, y, z) \tag{10}$$

The actual electric field strength is $2pT_g \nabla U(x,y,z)$.

The electric field and its components are shown in Figs. 5-7. Fig. 5 shows the 3D plots of $E_z(0,y,z)$ at several successive moments. It can be seen that $E_z(z)$ is symmetrical with respect to $z = 0$, and $E_z \to 0$ when $y$ and $z \to \infty$. In Fig. 6a, we show the planar sections of the 3D plots. The $z$-symmetry is clearly evident.

The magnitude of the $E_z$ amplitude ($\tilde{E}_z$) has been measured experimentally. The results are presented in Fig. 6b. Fig. 6b shows that $\tilde{E}_z(y)$ decays monotonically in the region $|z|<1$, while behaving non-monotonically in the region $|z|>1$.

The component $E_y(z)$ is an odd function of $z$, (see Fig. 7a). $E_y(z)$ has cusp shaped extremum points at $|z|=1$. The amplitude of $E_y(y)$, $\tilde{E}_y(y)$, Fig. 7b, decays monotonically with $y$ over a distance of ~$L$.

2.3. Equivalent circuit of the PE as a capacitive current source.

The thermal wave, propagating along the PE, generates in its vicinity an alternating electric field, and the associated displacement current. In fact, the PE acts as a capacitive source of alternating current. Two probe electrodes, connected to an electrometer, were placed in the region of the HEDEF. Since there is a potential difference between the two points where the probes are positioned, an alternating current will flow through the electrometer. The equivalent scheme of this experimental set-up is shown in Fig. 8.

The probe electrodes are attached to a dielectric plate to control their positioning. The displacement current, $I$, flowing from PE to the probe electrodes splits into the conductive current via electrometer ($I_r$), and the displacement current ($I_d$) through the probe electrode capacitance $C_d$. Then,

$$I = I_d + I_r;$$
$$\frac{I_d}{-i\omega C_d} = I_r R;$$
$$I_r = \frac{I}{1 + i\omega RC_d}; I_d = \frac{i\omega RC_d}{1 + i\omega RC_d} I; \quad (11)$$
$$I = -S\frac{i\omega}{4\pi} E(\vec{r})$$

Here, $\omega$ is the frequency of HEDEF, $S$ is the crossection area of the probe electrodes, $E(\vec{r})$ is the HEDEF at the probe electrode position. In our experiments $\omega RC_d \ll 1$, therefore $I_r \simeq I$.

It should be emphasized that a rigorous solution of the real system may be not only far too complicated but also impractical. Therefore we have chosen a simplified scheme of our system, neglecting the capacitive current via the probe electrodes.

3. Experiment

All the measurements were carried out on a PE sample placed in a cylindrical metal vacuum chamber. The polarization axis is directed along $z$ axis, i.e. the polar sides are $z = \pm L$ (*xy* planes). The HEDEF is measured in the region $y > 0$, i.e. at the non-polar face.

The experimental procedure consists of following steps:

1) The temperature distribution along the *y*-axis was determined under periodical variation of the temperature at the plane $y = -H$;

2) The spatial distribution of HEDEF was qualitatively determined by the method described in Section 2.3;

3) The dependences of the temperature distribution, and magnitude of HEDEF on air pressure, were determined;

4) The effect of covering the PE by Teflon sheets, to reduce the screening of the polarization charges by charged particles from the surrounding media, was studied qualitatively.

3.1. The thermal wave measurement

Eqs. (4) and (5) are required for the calculation of the EDEF [using Eq. (10)]. Thus the purpose of the thermal wave measurements was to test the validity of Eqs. (4) and (5), and to determine the values of the parameters $\chi$ and *b*.

Two thermocouples were used in these measurements (see Fig. 1b). One, differential, thermocouple, was measuring the difference *T* between the temperature at point $y = a$ (on one of its polar side faces) and the temperature $T_g = T(-H)$, at a point in immediate proximity of the thermoelectric-heater/PE interface. The second thermocouple was measuring the temperature $T_g$. Since the frequency *f* of the temperature alternation was very low, the periods were $\tau = 20 \div 250$ s, the temperature kinetics could be recorded digitally in great detail. Dependence on the ambient air pressure, in an interval of $\mathcal{P} = 760 \div 10^{-6}$ torr, was also measured.

The experimentally derived $T(\tau, \mathcal{P})$ together with Eq. (4) allows to determine [4] the values of both *b* and $\chi = 0.0115$ cm$^2$/s.

Fig. 9a demonstrates the excellent fitting of the experimental and calculated functions $T(\tau,\mathcal{P})/T_g$. Fig. 9b shows the function $b(\mathcal{P})$ derived from that fitting. As expected on physical grounds, $b(\mathcal{P})$ decreases with pressure, and saturates at high and low pressures. Thus, these measurements prove that the thermal model, described in part 2.1, is valid, and also provide the values of $\chi$ and $b(P)$, and, hence, the function $T(y)$ required to calculate HEDEF.

3.2. Space distribution of HEDEF, $E_z(y)$.

The formulas for HEDEF take into account only the surface polarization charge, and do not consider the charge adsorbed on the surface due to charged particles from the surrounding medium. Thus, as expected, the agreement between the experimental results and the theory was better when the measurements were carried out at the low pressure of $10^{-6}$ torr.

The spatial distribution of the $E_z(y)$ was measured as depicted in Fig. 1b. SrTiO$_3$ plate or a glass plate were used as holders of the probe electrodes. The probe holder plate was placed parallel to the PE face $y = 0$. The probe electrodes were positioned symmetrically with respect to the plane $z = 0$. The amplitude of the current, $\tilde{I}$, the amplitude of the temperature variation $\tilde{T}_g$ at $y = -H$ (close to the heater element), and the amplitude of the temperature difference variation, $\tilde{T}$ (at the point $y = a$) were measured simultaneously, point by point, as a function of time $t$, for a series of periods $\tau$ of the harmonical thermal wave. The probe holder was then moved along the $y$ axis ($y > 0$), and the procedure was repeated. The results were then used to calculate the amplitude of $I$, ($\tilde{I}$), vs. the amplitude of $T$, ($\tilde{T}$), for different periods $\tau$. From this plot, choosing a constant $\tilde{T}$, one determines $\tilde{I}$ as a function of $\tau$. Then, the HEDEF amplitude, $\tilde{E}_{z,\exp}(x,y,z)$ was calculated using Eq. (11).

The amplitude of the z-component of the field, $\tilde{E}_{z,\exp}(x,y,z)$ thus found, was then compared with the $\tilde{E}_{z,th}(x,y,z)$ calculated using Eq. (10) (for details see Appendix 2), using the values of $\chi$ and $b$ found as described above (see part 3.1). The results are shown in Fig. 10.

The distance between the probe electrodes, as shown in Fig. 10, is 7.8 mm > 2L. The experimentally observed dependence on y coincides nicely with the, *non-monotonic*, theoretically calculated curve. This characteristic detail has been indicated in Fig. 6b.

The points measured when the distance between the probe electrodes < 2L coincide satisfactorily with theoretical curves. The adjusting parameter for both groups of graphs is the area of probe electrodes, which is difficult to determine directly. The values $\tilde{E}_{z,\exp}(x,y,z)$ obtained by fitting are rather reasonable. Thus, theoretical predictions, as described in Part 2, are confirmed experimentally. It is worth mentioning, that in treating the experimental data, the three theoretically calculated entities, the thermal wave, the electric field HEDEF and the equivalent circuit analysis, are inherently interlinked.

3.3. Dependence of HEDEF on ambient air pressure

The experimental results presented in Section 3.1 clearly indicate the dependence of $T(y,t,\tau)$ on the ambient air pressure. Since, according to Eqs. (8) and (9), HEDEF is determined by $T(y,t,\tau)$, it must also depend on air pressure via the parameter $b$. Moreover, the air pressure affects HEDEF via the adsorption. Therefore, it is of interest to reveal and compare these two influences.

The pressure dependence for $\mathcal{P} \sim 10^{-6} \div 760$ torr has been measured as described in Section 3.2. The distance between the probe electrodes was 2.5 mm. The glass holder was positioned at the $y = 0$ plane. Thus, the amplitude of the electric field $\tilde{E}_z$ ($y = 1.1$ mm, $\mathcal{P}$) reflects the effect of the pressure.

The results are shown in Fig. 11. For comparison, one calculated curve (for $\tau = 250$ sec) is shown. Fig. 12 presents the calculated dependence of HEDEF on $b$. Thus, the calculation does not take into consideration the adsorption. On the other hand the experimental curves shown in Fig. 11 include both effects – the pressure dependence of $T(y,t,\tau)$ via $b$, and the adsorption. The apparent difference between the experimental graphs (Fig. 11) and those calculated (Fig. 12), indicates the important role of adsorption on the change of HEDEF with pressure.

The graphs in Fig. 11 show that at long $\tau$, the pressure effect is much stronger and becomes non-monotonic. The enhanced role of adsorption with increasing $\tau$ is obvious, since the density of adsorption charge increases with duration of adsorption.

The origin of the charged particles, adsorbed on the PE and screening its polarization charges, is not clear. In particular, charged particles could arise by surface ionization of neutral adsorbed molecules [5], due to the very high values of $E_y$ at the PE edges (Fig.7a).

The HEDEF is generated by the change of the net surface charge ($\sigma_{net}$), which is the difference between the polarization charge and the adsorbed surface charge. Then, HEDEF reflects a periodical change of the net surface charge induced by thermal wave. Its amplitude is

$$\Delta\sigma_{net}(y, \mathcal{P}) = pT(y) - q_a n_{a1}(y, \mathcal{P}) \qquad (12)$$

where $n_{a1}$ is the change in the adsorbing particles density, and $q_a$ is their charge. The quantity $pT(y)/q_a$ plays the role of the change of density of adsorption centers due to the thermal wave. Then, the density of polarization charge, in units of $q_a$, can be interpreted as the density of adsorption centers, and $pT(y)/q_a$ as its change. In these terms, the thermal wave changes the density of adsorption centers, which induces a change of their occupation at given $\mathcal{P}$ and $\tau$. It is possible that the HEDEF behavior in the pressure range $10^3 - 10^{-2}$ torr can be explained by a growth of $\Delta\sigma_{net}(y)$ with decreasing pressure $\mathcal{P}$. In the lower pressure region, below about $10^{-2}$ torr, the HEDEF decreases slowly. The origin of this behavior is not clear yet.

At long periods, with decreasing pressure, HEDEF increases by as much as a factor of 3, as shown in Fig. 11.

3.4. Shielding the polarization charge from adsorption screening

The decrease of the HEDEF due to the adsorption can be reduced by shielding the polarization charges at the PE polar faces. This can be realized by weakening the HEDEF at the polar faces, so that fewer charged particles will be collected there. As a result the HEDEF above the $y = 0$ plane will increase.

The experiment was carried out using the same experimental set-up as shown in Fig. 1b. Teflon plates, having a thickness of 7 mm, have been glued on the polar faces $z = \pm L$ (see insert in Fig. 13). The temperature and the HEDEF were then measured, as explained above, at two pressures, 760 torr and $5.5 \times 10^{-4}$ torr. As seen in Fig. 13, with Teflon shield, the HEDEF increases markedly,

4. Summary

In this work we measured and characterized the EDEF, excited by a harmonical thermal wave. We present experimental results obtained using a pyroelectric crystal LiTaO$_3$, as well as the theoretical analysis and description of the thermal wave and the induced harmonically varying EDEF. We have shown that an equivalent circuit of a pyroelectric capacitive current source is an adequate approximation of the real physical system. The effect of the air pressure, at the PE/air interface, on the EDEF was determined in the interval $10^3 - 10^{-6}$ torr. We found that EDEF increases significantly with decreasing air pressure, presumably due to diminishing of adsorption screening at the polar faces. Teflon plates, covering the polar faces, prevent accumulation of screening charged particles, resulting in a drastic increase of EDEF.

The characteristics of EDEF as disclosed in this study point to a large number of potential application in the areas of deposition methods, molecular structuring and arrangement, manipulation of charged or polarizable nano-particles, as well as surface physics and chemistry.

References

1. V. Sandomirsky, Y. Schlesinger, R. Levin, The edge electric field of a pyroelectric and its applications, J. Appl. Phys., **100**, 113722 (2006)

2. S.B. Lang, Regular bibliographic guide published in "Ferroelectrics".

3. S. Yamaguchi, Surface electric field of tourmaline, Appl. Phys. A **31**, 183-185 (1983)

4. In the range of small $\tau \sim 20 - 60$ s, the measured function $T_a(\tau, \mathcal{P})$ does not depend on the pressure $\mathcal{P}$ and thus does not depend on $b$. This is due to the fact that at small $\tau$ the thermal wave length is short, thus the boundary condition, Eq. (2), at $y = 0$, is superfluous, and the thermal wave propagates as in a semi-infinite sample. This is true at all air pressures. Thus, only one unknown, $\chi$, remains to be found, The value thus obtained is $\chi = 0.0115$ cm$^2$/s. The other unknown, $b(\mathcal{P})$, is found from the experimentally determined $T_a(\tau, \mathcal{P})$ in the region of long $\tau$.

5. W. Tornow, S. M. Shafroth, J. D. Brownridge, Evidence for neutron production in deuterium gas with a pyroelectric crystal without tip, J. Appl. Phys., **104**, 034905 (2008).

Figure captions.

Figure 1 (a) Schematic drawing of the lines-of-force of a PE block.

**P** is the polarization; DEF the internal field; EDEF the edge displacement electric field.

(b) The model system as used in the calculations. The bottom of the PE is heated harmonically by a thermoelectric heater. The side faces at $z = \pm L$ carry the polarization charges $\pm\sigma(y,t)$. The probe holder can be moved along the y-axis. The black arrowheads represent the measuring probes attached to the probe holder.

Figure 2 (a) 3D plot of the thermal wave, $T(y, t)/T_g$.

The vertical y axis is $T/T_g$ ; the time axis is in units of $\delta = \omega t$; the position axis is in cm.

$\tau = 100$ s; $b = 0.55$; $\chi = 0.0115$ cm$^2$/s.

(b) The cross-sections of the 3D plot of the thermal wave (Fig.2a) by planes $\delta$ = const.

$\tau = 100$ s, $b = 0.55$, $\chi = 0.0115$ cm$^2$/s.

Figure 3 The potential $U(y,z)$ at three successive instants of time (phase $\delta$).

(a) $\delta = 0$; (b) $\delta = 3\pi/4$; (c) $\delta = \pi$

Notice the change of scale in (b).

Figure 4 (a) The dependence of the amplitude $\tilde{U}$ of the potential on period $\tau$ at the points (x,y,z): (0, 0.01, 1), (0, 0.01, 0.5), (0, 0.01, 0.25)

(b) The amplitude $\tilde{U}$ (z) of the potential for different values of the period $\tau$. $b = 0.55$; $x = 0$; $y = 0.01$

Figure 5 3D plots of $E_z$ for different $\omega t$: (a) 0, (b) $3\pi/4$, (c)$\pi$.

Notice the change of scale in (b).

Figure 6  (a) The profiles of $E_z(z)$ at different values of $\omega t$. $\tau = 100$ s; $b = 0.55$.

(b) The amplitude $\tilde{E}_z(y)$ at different $z$. $\tau = 100$ s; $b = 0.55$.

Figure 7  (a) The profiles of $E_y(z)$ at different values of $\omega t$.

$\tau = 100$ s; $b = 0.55$, y = 0.01.

(b) The amplitude $\tilde{E}_y(y)$ at different $z$.

$\tau = 100$ s; $b = 0.55$.

Figure 8  The equivalent circuit.

$R$ is the resistance of the electrometer branch. The probe electrodes (PR) are connected by $C_d$ - the mutual capacitance. $C_0$ and $C_1$ are equivalent capacitances.

Figure 9  The characteristics of the thermal wave.

(a) $\tilde{T}(\tau, \mathcal{P})$ – experimental (symbols) and calculated (solid) graphs.

$\mathcal{P}$: (1) $10^{-5}$ torr; (2) $10^{-3}$ torr; (3) $10^{-1}$ torr; (4) 760 torr.

(b) The function $b(\mathcal{P})$.

Figure 10  The dependence of the EDEF amplitude $\tilde{E}_z(y, \tau)$ on $y$.

Distance between probes 15.6 mm (beyond PE edges); STO-holder; $\mathcal{P} = 760$ torr; $\tilde{T}_g = 0.3$ K.

The solid curves are calculated using Eqs.(9, 10). The adjusting parameter is the area of probe electrodes, $S = 1.7$ mm$^2$.

$\tau$: (a) 40 sec; (b) 100 sec; (c) 250 sec.

Figure 11  The pressure dependence of the HEDEF amplitude at y = 1.1 mm

Distance between probes 2.5 mm; glass holder.

τ: (1) 250 sec; (2)160 sec; (3) 80 sec; (4) 50 sec; (5) 26 sec; (6) 18 sec.

Curve (7) is calculated using Eq. (10) for τ = 250 sec. Notice that in the calculations only the dependence of $b(\mathcal{P})$ on the temperature has been accounted for, i.e. the effect of adsorption was neglected.

Figure 12    Theoretical pressure dependence (via $b$) of the HEDEF amplitude.

The vertical axis in units of $E_0 = 2pT_g = 13.25$ kV/cm.

τ: (1) 240 sec; (2)120 sec; (3) 60 sec; (4)30 sec; (5)10 sec.

Figure 13    The HEDEF amplitude $\tilde{E}_z$, at $y = 1.1$ mm, with Teflon shield.
Distance between probes 2.5 mm; glass holder.
$\mathcal{P}$: (1a) $2 \cdot 10^{-5}$ torr; (1b) $5 \cdot 10^{-4}$ torr; (2a) 760 torr.

Appendix 1. The thermal wave in the PE-bar

Let $T(y)$ be the temperature of PE at height $y$, relative to $T_g$, and $T_a$ the amplitude of temperature variation in the air. Then, neglecting heat losses at the *xy* and *zy* surfaces, as justified by the experimental results, the thermal wave behavior is determined by:

$$\frac{\partial T}{\partial t} = \chi \frac{\partial^2 T}{\partial y^2}; \quad t > 0; \; -H < y \leq 0$$

$$\frac{\partial T_a}{\partial t} = \chi_a \frac{\partial^2 T_a}{\partial y^2}; \; t > 0; \quad y > 0$$

A1.1

$$T(-H) = T_g \exp(-i\omega t); \; T(0) = T_a(0);$$

$$\kappa \left.\frac{\partial T}{\partial y}\right|_0 = \kappa_a \left.\frac{\partial T_a}{\partial y}\right|_0; \; T_a(\infty) = 0$$

A1.1.1

The dependence of $T$ on $y$ can then be expressed as:

$$T = C_1 \exp(ky) + C_2 \exp(-ky) \quad \text{A1.2.1}$$

and

$$T_a = C_3 \exp(-k_a y) \quad \text{A1.2.2}$$

Applying the boundary condition

$$T_g = C_1 \exp(-kH) + C_2 \exp(kH) \quad \text{A1.3.1}$$

and the boundary conditions given above (A.1.1.1), one obtains:

$$C_1 + C_2 = C_3 \qquad \text{A1.3.2}$$

$$C_1 - C_2 = -b \cdot C_3 \qquad \text{A1.3.3}$$

$$k_0 = \sqrt{\frac{\omega}{2\chi}} = \sqrt{\frac{\pi}{\chi\tau}};\ k = (1-i)k_0; \qquad \text{A1.4.1}$$

$$k_{a0} = \sqrt{\frac{\omega}{2\chi_a}} = \sqrt{\frac{\pi}{\chi_a\tau}};\ k_a = (1-i)k_{a0} \qquad \text{A1.4.2}$$

$$b = \frac{\kappa_a k_a}{\kappa k} = \frac{\kappa_a}{\kappa}\frac{k_{a0}}{k_0} = \frac{\kappa_a}{\kappa}\sqrt{\frac{\chi}{\chi_a}} = \sqrt{\frac{\kappa_a c_a \rho_a}{\kappa c \rho}}. \qquad \text{A1.4.3}$$

where $c_a$ and $\rho_a$ are the heat capacity and density of air; $c$ and $\rho$ are the heat capacity and density of PE. Using the above expressions one obtains:

$$C_1 = \frac{(1-b)T_g}{2[\cosh(kH) + b\sinh(kH)]} \qquad \text{A1.5.1}$$

$$C_2 = \frac{(1+b)T_g}{2[\cosh(kH) + b\sinh(kH)]} \qquad \text{A1.5.2}$$

$$C_3 = \frac{T_g}{\cosh(kH) + b\cdot\sinh(kH)} \qquad \text{A1.5.3}$$

The solution of Eq. (A.1.1) is then:

$$T(y,t) = T_g \exp(-i\omega t)\cdot\frac{\cosh(ky) - b\sinh(ky)}{\cosh(kH) + b\sinh(kH)} \qquad \text{A1.6}$$

For the purpose of comparison with the experiment, the magnitude of the temperature variation amplitude is then:

$$\tilde{T}(y) = T_g \frac{\left[(\cosh\alpha - b\sinh\alpha)^2 \cos^2\alpha + (b\cosh\alpha - \sinh\alpha)^2 \sin^2\alpha\right]^{1/2}}{\left[(\cosh\beta + b\sinh\beta)^2 \cos^2\beta + (b\cosh\beta + \sinh\beta)^2 \sin^2\beta\right]^{1/2}} \qquad \text{A1.6.1}$$

Appendix 2. HEDEF of the PE-bar

As described in Section 1.1, the thermal wave generates harmonically varying polarization charges at the faces $z = \pm L$. Using Eq. (7), $\Delta\sigma(y) = pT(y)$, it is straightforward to calculate the potential due to the varying polarization charges.

As derived in Appendix 1, $T(y)$ in the region $-H < y < 0$ is given by

$$T(y,t) = T_g \exp(-i\omega t)\frac{\cosh(ky) - b\sinh(ky)}{\cosh(kH) + b\sinh(kH)} \tag{A2.1}$$

Let the width of the PE be $2L$, $(-L < z < L)$, and its thickness $2W$, $(-W < x < W)$, then, the potential will be

$$\Phi(x,y,z) = \Phi_1(x,y,z) + \Phi_2(x,y,z) \tag{A2.2}$$

$$\Phi_1(x,y,z) = 2\int_0^W du \int_{-H}^0 dv \frac{-pT(v,t)}{\sqrt{(x-u)^2 + (y-v)^2 + (z+L)^2}} = \tag{A2.2.1}$$

$$= \frac{-2pT(-H)\exp(-i\omega t)}{\cosh(kH) + b\sinh(kH)} \int_0^W du \int_{-H}^0 dv \frac{\cosh(kv) - b\sinh(kv)}{\sqrt{(x-u)^2 + (y-v)^2 + (z+L)^2}}$$

$$\Phi_2(x,y,z) = 2\int_0^W du \int_{-H}^0 dv \frac{pT(v,t)}{\sqrt{(x-u)^2 + (y-v)^2 + (z-L)^2}} = \tag{A2.2.2}$$

$$= \frac{2pT(-H)\exp(-i\omega t)}{\cosh(kH) + b\cdot\sinh(kH)} \int_0^W du \int_{-H}^0 dv \frac{\cosh(kv) - b\cdot\sinh(kv)}{\sqrt{(x-u)^2 + (y-v)^2 + (z-L)^2}}$$

Integrating over $u$ results in:

$$I_1(v,x,y,z) = \int_0^W du \frac{1}{\sqrt{(x-u)^2 + (y-v)^2 + (z+L)^2}} =$$
$$= \ln\left|\frac{(x-W) + \sqrt{(x-W)^2 + (y-v)^2 + (z+L)^2}}{x + \sqrt{x^2 + (y-v)^2 + (z+L)^2}}\right| \tag{A2.3.1}$$

$$I_2(v,x,y,z) = \int_0^W du \frac{1}{\sqrt{(x-u)^2 + (y-v)^2 + (z-L)^2}} =$$
$$= \ln\left|\frac{(x-W) + \sqrt{(x-W)^2 + (y-v)^2 + (z-L)^2}}{x + \sqrt{x^2 + (y-v)^2 + (z-L)^2}}\right| \quad (A2.3.2)$$

Then, inserting Eqs. A2.3 into Eqs. A2.2 respectively, one obtains:

$$\Phi_1(x,y,z) =$$
$$= \frac{-2pT(-H)\exp(-i\omega t)}{\cosh(kH) + b\sinh(kH)} \int_{-H}^0 dv[\cosh(kv) - b\sinh(kv)]I_1(v,x,y,z) \quad (A2.4.1)$$

$$\Phi_2(x,y,z) =$$
$$= \frac{2pT(-H)\exp(-i\omega t)}{\cosh(kH) + b\sinh(kH)} \int_{-H}^0 dv[\cosh(kv) - b\sinh(kv)]I_2(v,x,y,z) \quad (A2.4.2)$$

Then the dimensionless potential will be given by:

$$U(x,y,z) = \frac{\Phi_1 + \Phi_2}{2pT(-H)} \quad (A2.5)$$

A tedious, but straightforward calculation results in the following expression for the potential $U$:

$$U =$$
$$= \frac{\cos\omega t - i\sin\omega t}{D_1^2 + D_2^2} \int_{-h}^0 dv[C_1(\beta,\eta) + iC_2(\beta,\eta)][-I_1(v,x,y,z) + I_2(v,x,y,z)] \quad (A2.6)$$

Where

$$D_1(\beta) = [\cosh\beta + b\sinh\beta]\cos\beta; \quad D_2(\beta) = [\sinh\beta + b\cosh\beta]\sin\beta \quad (A2.6.1)$$

and $C_1$ and $C_2$ are defined by Eqs. A1.5.1 and A1.5.2.

The amplitude $\tilde{U}(x,y,z)$ is then:

$$\tilde{U}(x,y,z) = \frac{1}{\sqrt{D_1^2 + D_2^2}}\left\{\left[\int_{-h}^0 dvC_1(-I_1 + I_2)\right]^2 + \left[\int_{-h}^0 dvC_2(-I_1 + I_2)\right]^2\right\}^{1/2} \quad (A2.7)$$

The dimensionless electric field is then found from Eq. (10):

$$\vec{E}(x,y,z) = \nabla_{x,y,z}[\operatorname{Re} U(x,y,z)] =$$

$$= \frac{\cos\delta}{D}\int_{-h}^{0} dv\, C_1(-\nabla I_1 + \nabla I_2) + \frac{\sin\delta}{D}\int_{-h}^{0} dv\, C_2(-\nabla I_1 + \nabla I_2) \qquad (A2.8)$$

and the electric field (expressed in units of V/cm) is obtained by:

$$2pT(-H)\vec{E}(x,y,z). \qquad (A2.8.1)$$

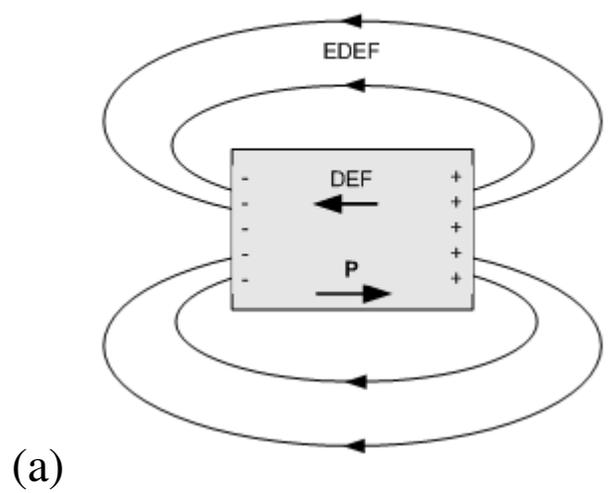

(a)

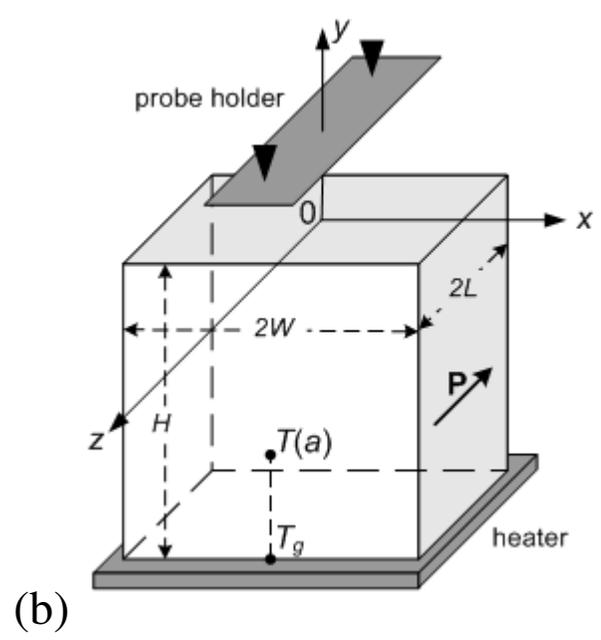

(b)

Fig. 1

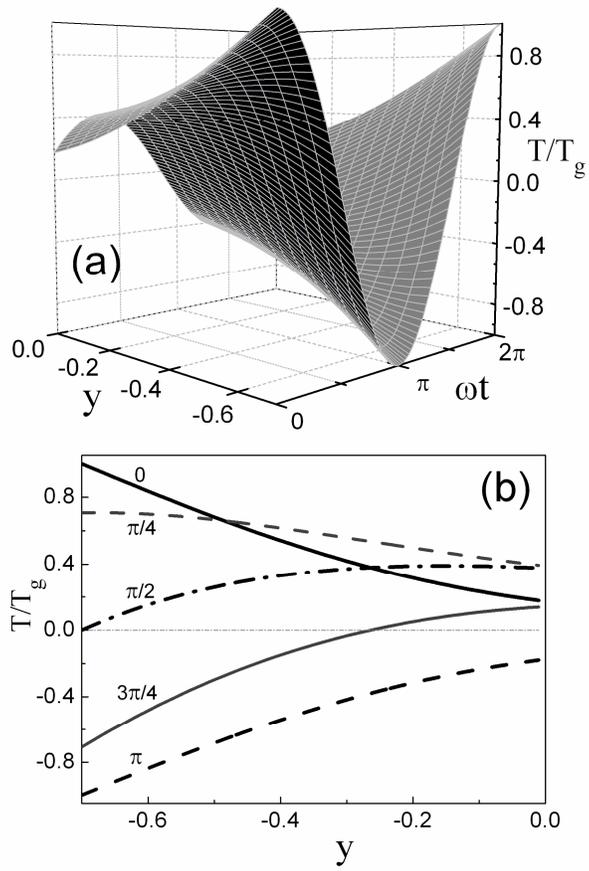

Fig. 2

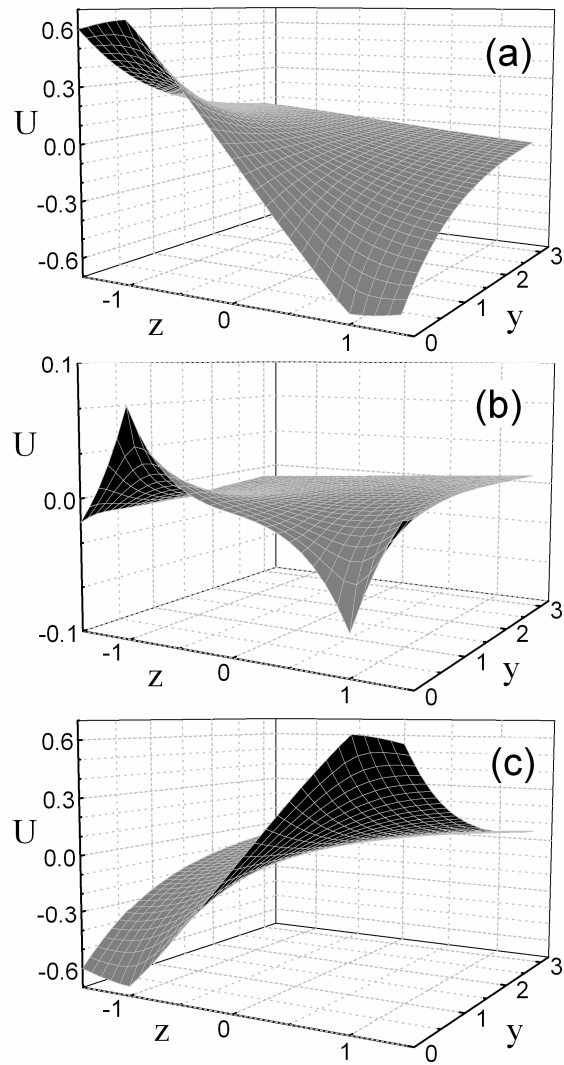

Fig. 3

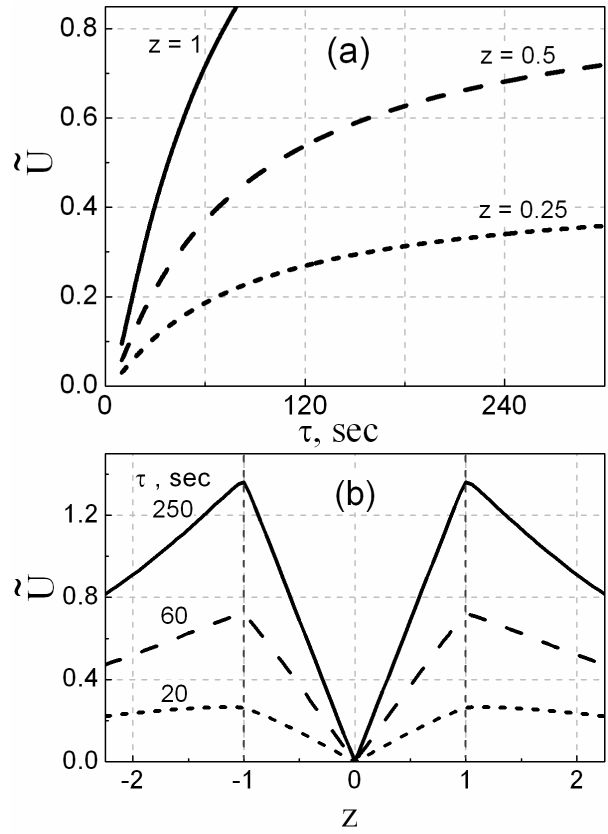

Fig. 4

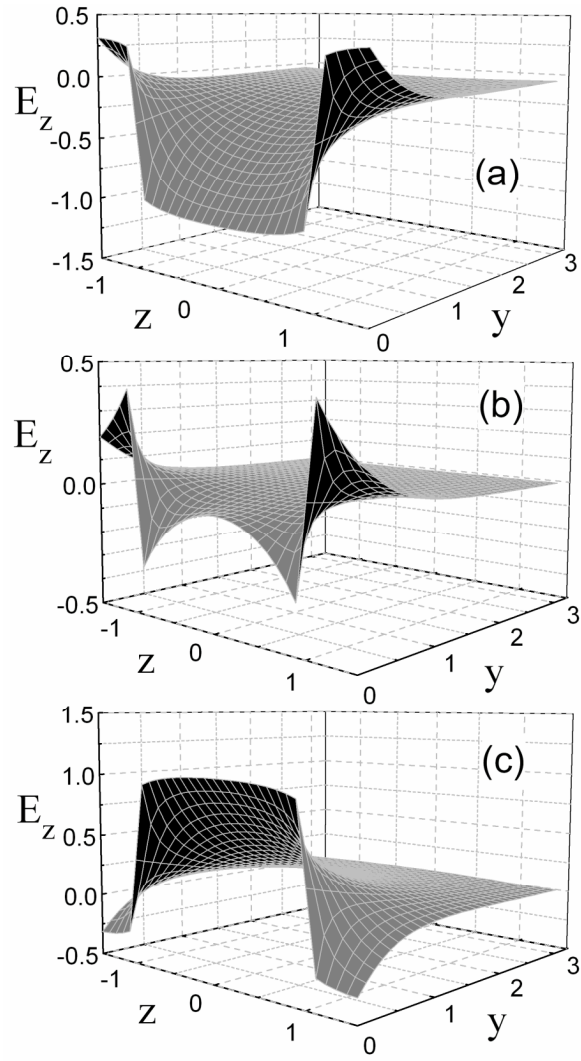

Fig. 5

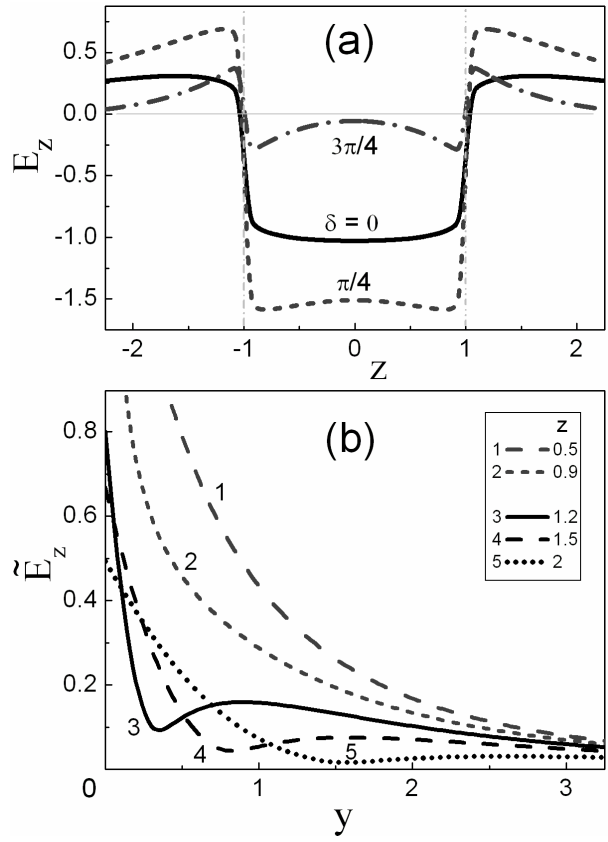

Fig. 6

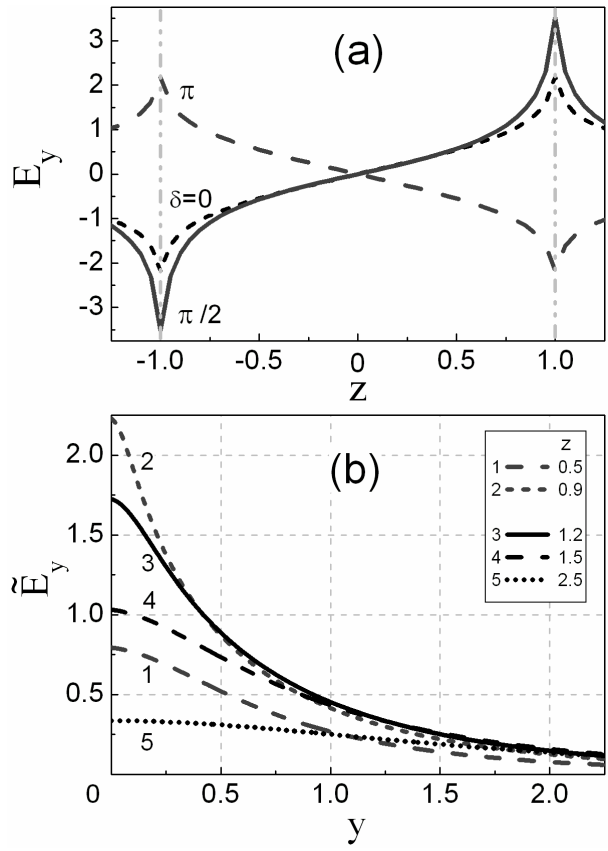

Fig. 7

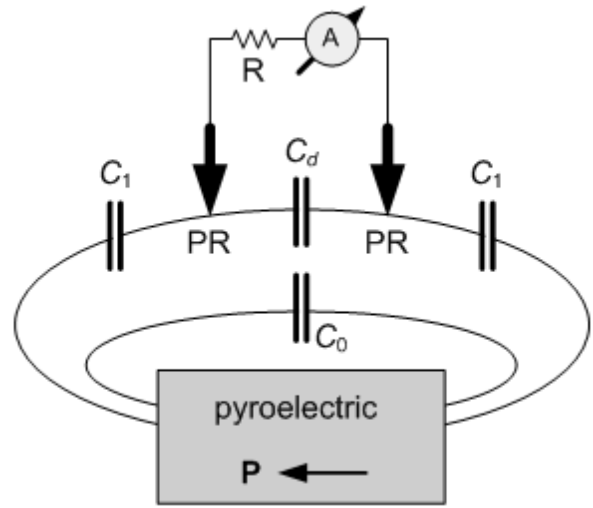

Fig. 8.

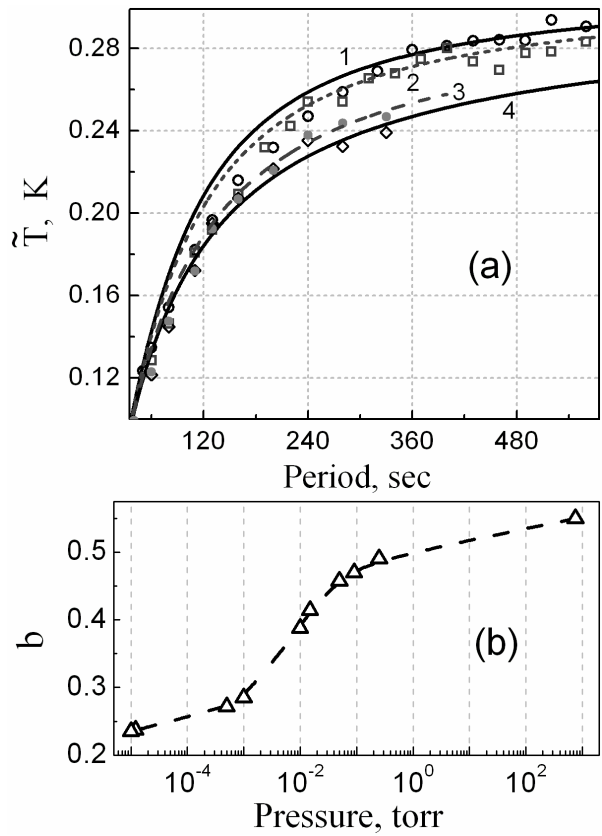

Fig. 9

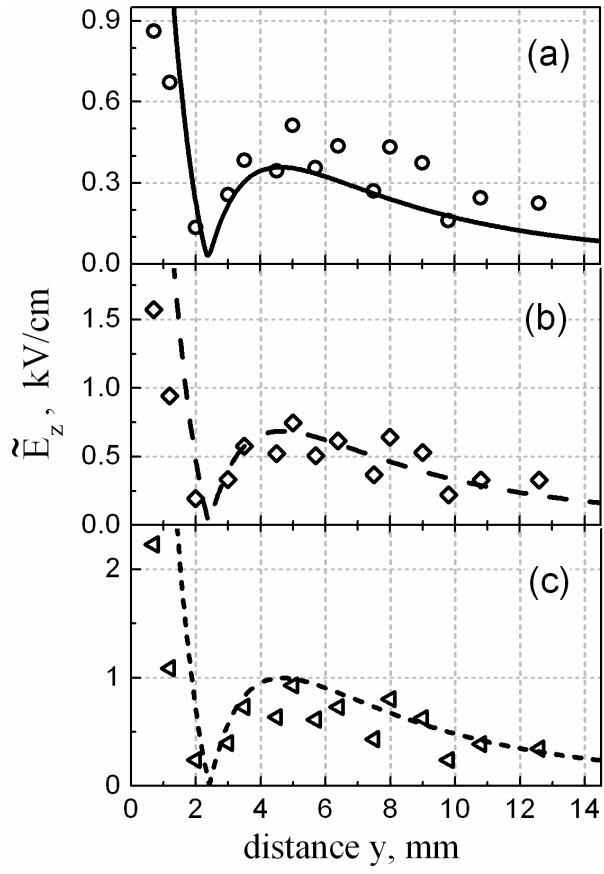

Fig. 10

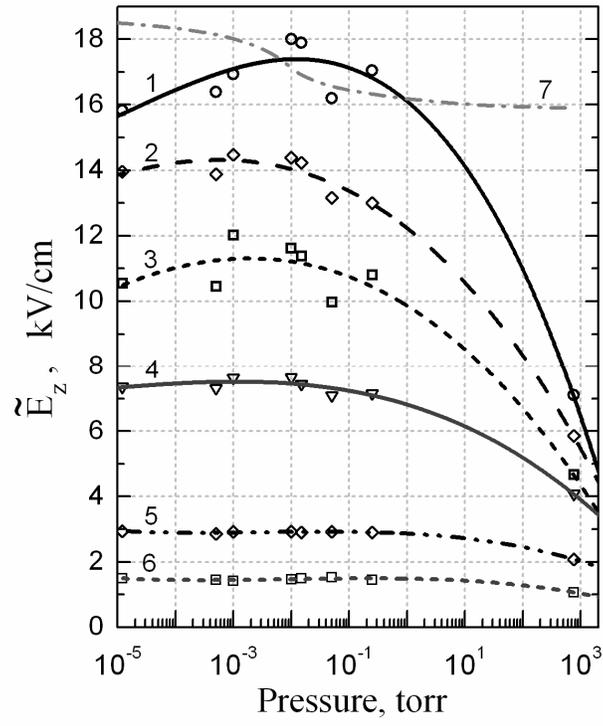

Fig. 11

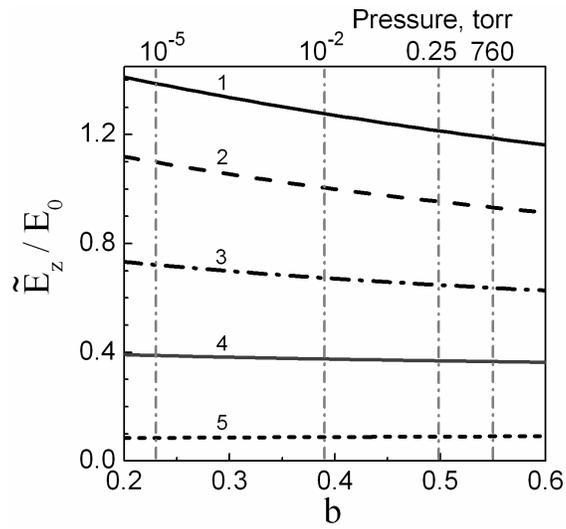

Fig. 12

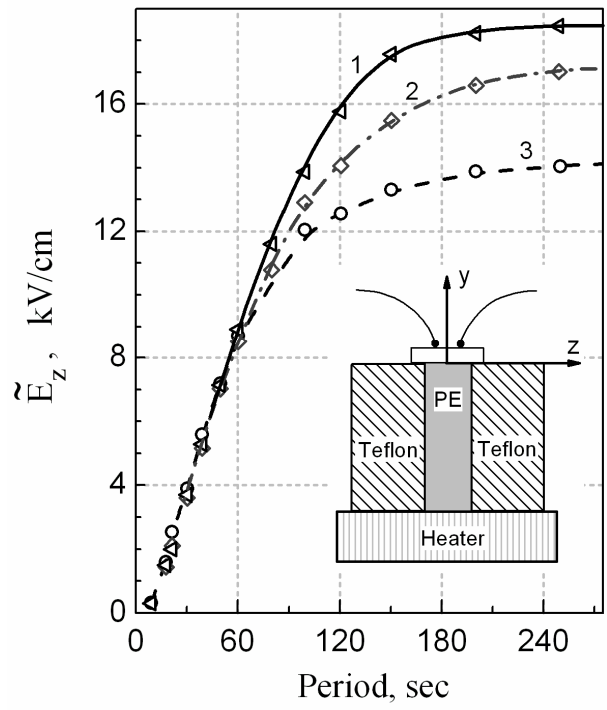

Fig. 13